# Deep Learning Based Solar Cell Recognition for Optical Wireless Power Transfer

Sida Huang[1] and Yuanting Wu[1] and Dinh Hoa Nguyen[1,2*]
[1]Graduate School of Mathematics, Kyushu University, Fukuoka, Japan
[2]International Institute for Carbon-Neutral Energy Research (I2CNER), and Institute of Mathematics for Industry (IMI), Kyushu University, Fukuoka, Japan. Email: hoa.nd@i2cner.kyushu-u.ac.jp
[*]Corresponding author

**Abstract**

Optical wireless power transfer (OWPT) is a technology that wirelessly transmit light energy from an optical transmitter to an optical receiver, usually a solar cell. In order to achieve the highest transmission efficiency, the solar cell receiver should be accurately aligned with the optical transmitter. Hitherto, only a few works have been existed for solar cell recognition in presence of complex backgrounds. In this paper, we employ a deep learning approach based on Yolov5-Lite for the solar cell recognition purpose, due to its lightweight, fast and easy to deploy on hardware characteristics. Our tests show a high accuracy of the employed deep learning model with the highest F1 score of 91% and mAP of 94.8%. Therefore, this deep learning model is highly promising for use in OWPT systems to precisely align optical transmitters and solar cell receivers.

## 1. INTRODUCTION

WPT is a method of transmitting power without physical connections between transmitters and receivers. Because of its flexibility, convenience, and other advantages, WPT has been widely used in many fields such as electric vehicles (EVs) and consumer electronics. There are three mainstream WPT technologies: magnetic coupling WPT, microwave WPT, and optical WPT (OWPT) [1]. Among them, magnetic coupling WPT faces the problem of short transmission distances, whereas microwave WPT suffers from the impact of electromagnetic waves [2]. In contrast, OWPT is a highly promising method for long distance power transmission and no electromagnetic interferences. Therefore, OWPT has been considered as a great solution for many areas, such as satellites [3], medical implants [4], electric vehicles [5], and so forth.

As depicted in Figure 1, an OWPT system consists of three basic parts: an optical transmitter, a transmission environment, and an optical receiver [6]. The light source can be a laser diode (LD) or a light emitting diode (LED), while the light receiver can be a solar cell or a photodiode employed to convert optical signals into electric power. As such, OWPT systems have relatively simple and compact structures [7], compared to other WPT technologies mentioned above.

In order to convert light energy into electrical energy with high efficiency, light beams must be precisely projected onto the optical receiver. Hence, the accurate positioning of the receiver is essential. As artificial intelligence (AI) technologies advance rapidly, they bring innovative solutions to various problems. This study employs the YOLOv5-Lite, a deep learning model, for the purpose of precisely detecting solar cell optical receivers.

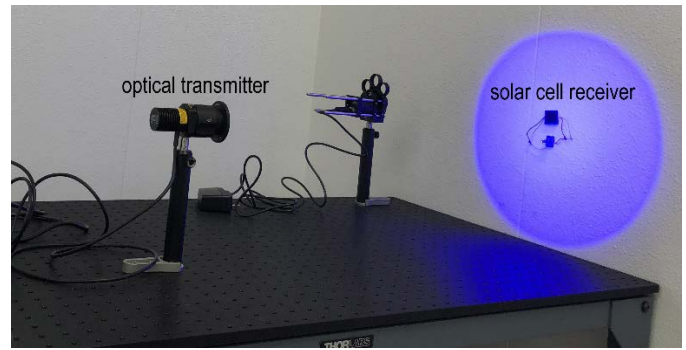

Figure 1. Illustration of an OWPT system.

## 2. BACKGROUND

The energy conversion efficiency of an OWPT system relies on the capability of aligning the optical transmitter and receiver accurately. Hence, there have been several studies in the literature for the detection of solar cells. Kawashima et al. placed a corner cube placed in the center of the solar panel and detected its reflected light by a 4-element diode through a lens [8]. Wang X et al. and Lim et al. used retro-directive mirrors for solar cell detection [9], [10]. Putra et al. put an infrared LED marker on the target, then used OpenCV to process the image from a web camera for target recognition [1]. Sovetkin et al. proposed a method based on specialized Hough transforms to extract the solar cells even when they are surrounded by disturbing background [11]. Tang Jing et al. trained a classifier file provided by OpenCV to recognize solar cells [2]. Kaoru Asaba et al. utilized a differential absorption image sensor to achieve the purpose [12].

Among these existing studies, some have explored the installation of auxiliary devices (such as cube and LED marker) on solar cells for detection purposes. However, this approach increases the cost of identification, which is not conducive to large-scale deployments. Moreover, only a few studies have addressed the recognition of solar cells in the presence of complex backgrounds. However, the widespread adoption of OWPT is bound to encounter numerous complex application environments. Particularly in the case of dynamic OWPT (DOWPT), there are high demands for the accuracy and speed of identifying moving solar cell receivers. In addition, most of the existing researches have employed traditional image processing methods, which exhibit poor anti-interference capabilities, slow response times, and susceptibility to background interferences. In contrast to traditional image processing, deep learning methods offer numerous advantages by simulating human learning processes.

One significant advantage of deep learning in image processing is the reduced reliance on feature engineering (FE) [13]. Traditionally, image recognition tasks heavily depend on hand-engineered features, which significantly affected to the overall performance. For instance, detecting solar panels using conventional methods might require FE components based on the Hough transform for extracting edge features. FE is a complex and time-consuming process that needs adjustment whenever the problem or dataset changes, making it expensive and not well generalizable. Furthermore, manual search for effective FE is not an easy and obvious task. In contrast, deep learning eliminates the need for FE by autonomously identifying crucial features through training. Additionally, deep learning models demonstrate strong generalization capabilities. Deep learning models exhibit robustness under challenging conditions, including variations in illumination, complex backgrounds, different resolutions, sizes, and orientations of the images [14]. On the other hand, although deep learning typically requires longer training time compared to other traditional approaches like SVM, its efficiency during testing is notably faster. For example, in tasks such as obstacle detection and anomaly identification, deep learning models may take longer to train initially but exhibit shorter testing time compared to SVM and KNN [15]. Besides, given the time required for manual filter design and FE, the time for annotating images and training the deep learning model could be almost negligible [16].

To this end, the current research adopts deep learning methods and uses the YOLO series algorithm, which has made great achievements in the field of object detection, to recognize solar cells. This will enrich the application of deep learning in OWPT, and hopefully enable OWPT to better adapt for more complex scenarios in realistic applications.

## 3. METHOD

### 3.1. Deep Learning Approach for Image Recognition

Object recognition algorithms mainly compose of two-stage and one-stage algorithms. Two-stage algorithms, such as region-based convolutional neural networks (R-CNN) series algorithms, first generate region proposals which are sub-regions that may contain a target, and then proceed to perform classification and regression on these proposals. On the other hand, one-stage algorithms, such as the YOLO series algorithms, utilize a single CNN operation to directly predict the categories and positions of various targets. The R-CNN series algorithms excel in target detection scenarios that demand higher precision, but their detection speed is lower than that of the YOLO series algorithms [17]. In the context of OWPT (especially DOWPT), we aim for algorithms with superior real-time detection performance, and hence, YOLO series algorithms are adopted. These algorithms utilize regression models, making it easier to learn generalized features of targets and achieve recognition at a faster pace.

YOLO algorithms divide images into $S \times S$ grids for detection. Each grid cell is responsible for detecting objects if the center of an object falls within it. Additionally, each grid cell predicts $B$ bounding boxes, confidence scores, and $C$ class probabilities [18]. These bounding boxes represent the object's size and location in the image, while confidence scores indicate the model's certainty about object presence and prediction accuracy. As depicted in Figure 2, the YOLO network structure consists of two fully connected layers and 24 convolution layers. Following the fully connected layer, an output tensor of $S \times S \times (B * 5 + C)$ is generated. The final detection results are obtained by regressing the box positions and assessing the class probabilities of the tensor data [19].

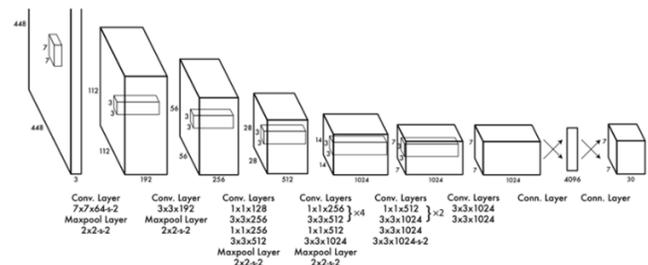

Figure 2. YOLO network structure.

In many OWPT systems, solar cells acting as receivers are often small, especially when viewed from the perspective of the light source over long distances. Therefore, robust detection of small targets, i.e. solar cells, is crucial. Although YOLO algorithms can achieve a rapid detection of targets, its performance in detecting small targets is suboptimal due to the lack of a detailed grid division, leading to multiple targets being assigned to the same grid [20], [21]. As one of the YOLO series algorithms, YOLOv5 [22] can overcome this issue. YOLOv5 improves training data quality by passing each batch through a data loader, which applies three types of data augmentation: scaling, color space adjustment, and mosaic enhancement. Additionally, the anchor mechanism of faster R-CNN is utilized to enhance YOLOv5's capability to detect small targets. This approach also enhances the algorithm's versatility across images of varying sizes.

In the system design of OWPT, additional programs are often required to control the movement of the light beam [7]. However, hardware memory resources are limited. To ensure the integration of various functional programs in OWPT system hardware, the memory consumption of the programs needs to be considered when selecting the receiver recognition method. YOLOv5-Lite [23] is a lightweight design of YOLOv5. The authors conducted a series of relaxation experiments on YOLOv5. By reducing the model's flops, lowering memory consumption, and minimizing parameters, the algorithm becomes lighter. Adding shuffle channels and reducing YOLOv5 head for channels make it faster. By removing the focus layer and four slice operations and adjusting the model's quantization accuracy to an acceptable range, it becomes easier to deploy. Therefore, this study chooses YOLOv5-Lite for solar cell recognition due to its high accuracy, lightweight design, fast speed, and ease of deployment.

### 3.2. Dataset and Processing

Due to the lack of publicly available standardized datasets for OWPT solar cells, we create our own datasets using four different types of solar cells with different sizes, shape and colors. To ensure dataset diversity, solar cell images are captured from various angles at different shooting distances under multiple backgrounds at different times. A total of 358 solar cells images with a resolution of 4032×3024 and JPG format was obtained. To create more complex background scenarios, solar cells are placed alongside various other objects, particularly those with similar appearances such as phones and iPads, as depicted in Figure 3. The dataset includes reflections, blurs, shadows, occlusions, low light conditions, overlapping solar cells, etc.

YOLOv5 officially recommends a dataset management platform called Roboflow [24]. Researchers can upload their datasets and perform operations such as format conversion, annotation, and label export. Additionally, various data preprocessing and augmentation functions are provided to help researchers optimize their data. In this study, images are randomly divided into 80% training set and 20% test set. Roboflow is used to draw external bounding boxes around solar cell targets, enabling manual annotation of solar cells. Images are annotated based on the minimum rectangle surrounding the solar cells to ensure that the rectangle contains as little background area as possible. The annotations are then saved and exported as TXT format files for training purposes. Additionally, to improve the efficiency of model reading and training data, preprocessing is conducted on the dataset, including auto-orientation and resizing. EXIF rotations are discarded, and pixel ordering is standardized. Images are resized to 640 x 640 pixels to obtain smaller file sizes and faster training.

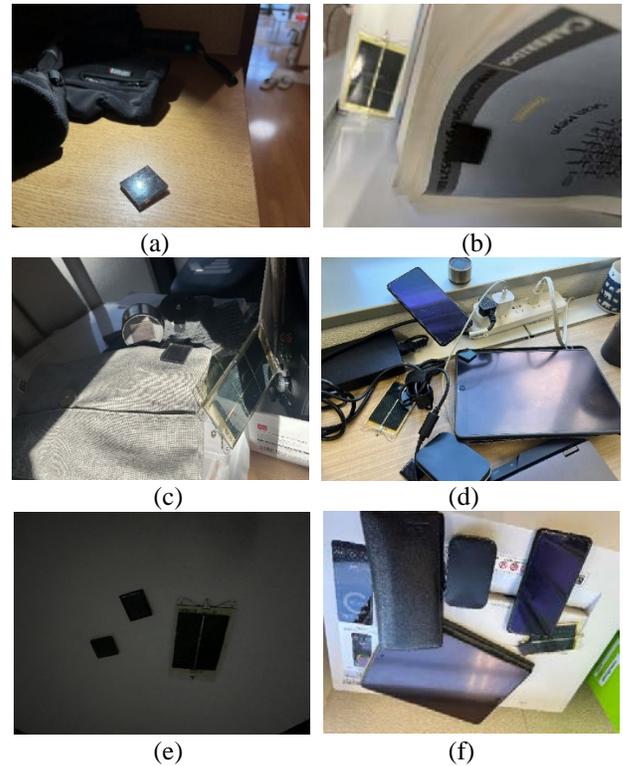

Figure 3. Solar cells images in different conditions: (a) reflections; (b) blurs; (c) shadows; (d) occlusions; (e) low light conditions; and (f) overlapping solar cells.

The generation of deep learning-based object detection models requires a large amount of data. Due to the limited number of samples, in order to better extract features of solar cells under complex backgrounds and avoid overfitting of the trained model, data augmentation is necessary. In this study, we employ various methods of data augmentation using Roboflow on the training set, increasing the number of training images from 286 to 804. Data augmentation operations include crop, rotation, shear, grayscale, hue, saturation, brightness, exposure, blur, noise, cutout, etc., as demonstrated in Figure 4.

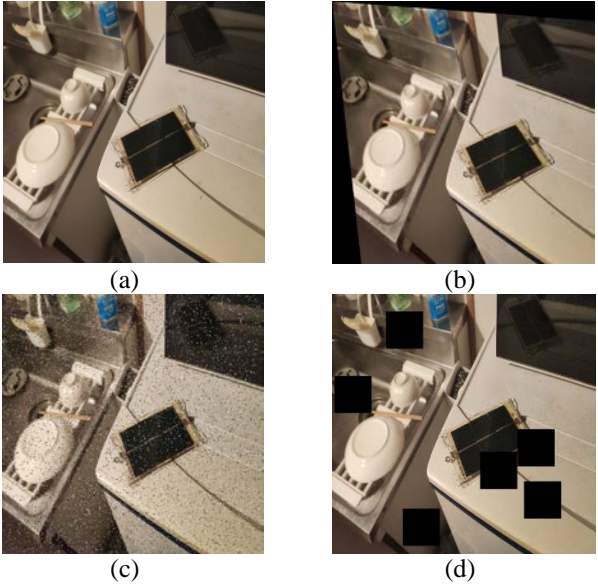

Figure 4. Different Data Augmentation:
(a) origin; (b) shear; (c) noise; and (d) cutout.

### 3.3. Experimental Platform

The training of the YOLOv5-Lite is conducted on the Windows 11 operating system. The CPU is 12th Gen Intel(R) Core(TM) i7-12700H CPU 2.7GHz with 16GB memory, the GPU is NVIDIA Geforce RTX 3060 Laptop GPU with 6GB video memory. The software environment is Cuda 11.6 and cuDNN 8.8.0. The deep learning framework is PyTorch 1.13.1.

### 3.4. Model Evaluation Indicators

In this research, objective evaluation metrics including Precision (P), Recall (R), mean average precision (mAP), and F1 score are employed to assess the effectiveness of the trained YOLOv5-Lite model. The formulas for Precision (P), Recall (R), and F1 score are as follows:

$$P = \frac{TP}{TP+FP} \quad (1)$$

$$R = \frac{TP}{TP+FN} \quad (2)$$

$$F1\ Score = \frac{2*P*R}{P+R} \quad (3)$$

True positives (TP), false positives (FP), and false negatives (FN) represent positive samples correctly classified, negative samples incorrectly classified as positive, and positive samples incorrectly classified as negative, respectively.

Average Precision (AP) is the integral of the precision-recall (P-R) curve, representing the average accuracy rate. Mean Average Precision (mAP) is the average of the AP values across all categories. Specifically, mAP is calculated by summing the AP values of each category and then dividing by the total number of categories, yielding the average value. The formulas for AP and mAP are as follows:

$$AP = \int_0^1 P(R)\, dR \quad (4)$$

$$mAP = \frac{1}{|Q_R|}\sum_{q=Q_R} AP(q) \quad (5)$$

where $P(R)$ is the P-R curve and $Q_R$ is the number of categories. mAP is used to measure the quality of the detection model. A higher mAP indicates higher average detection accuracy and better performance.

## 4. RESULT

In this paper, we configure the batch size to 8 and train the model for 100 epochs to obtain the results based on training set. The training results are shown in Figure 5.

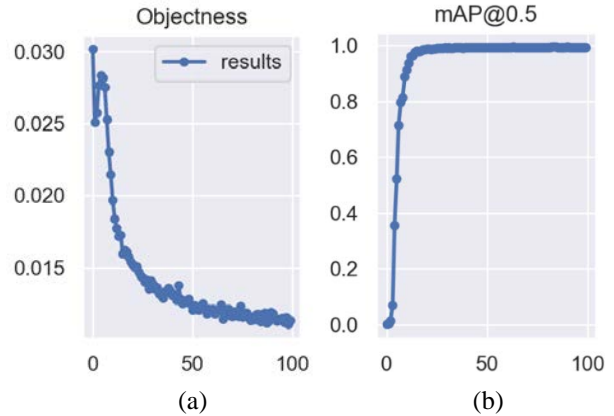

Figure 5. Training result.

In Figure 5(a), "objectness" refers to objectness loss, measures the confidence of the model in predicting the presence of an object within a bounding box. A smaller objectness loss indicates better target detection accuracy. According to Figure 5(a), it is visually observed that the YOLOv5-Lite network model's loss decreases as the number of iterations increases, and the loss curve gradually stabilizes and converges near 0. Figure 5(b) shows that after 100 iterations, the mAP of the YOLOv5-Lite model reaches a maximum of 99.4%, indicating high average accuracy for solar cell detection.

To further validate the model's performance in recognizing solar cells, we conduct the trained model on the test set, yielding results as shown in Figure 6. P and R scores (equations (1) and (2)) are typically conflicting performance metrics. Generally, when P score is higher, R score tends to be lower. To comprehensively consider these two metrics, we evaluate using F1 score and mAP. It can be observed from Figure 6(a) that the F1 score of the trained YOLOv5-Lite model reaches 91%, which is relatively high in object detection tasks. The mAP reaches 94.8% as shown in Figure 6(b), although it decreases compared to the training set, it still maintains a high level of accuracy. Additionally, the model's inference time per image is 9.4 ms, meeting the requirements for real-time recognition.

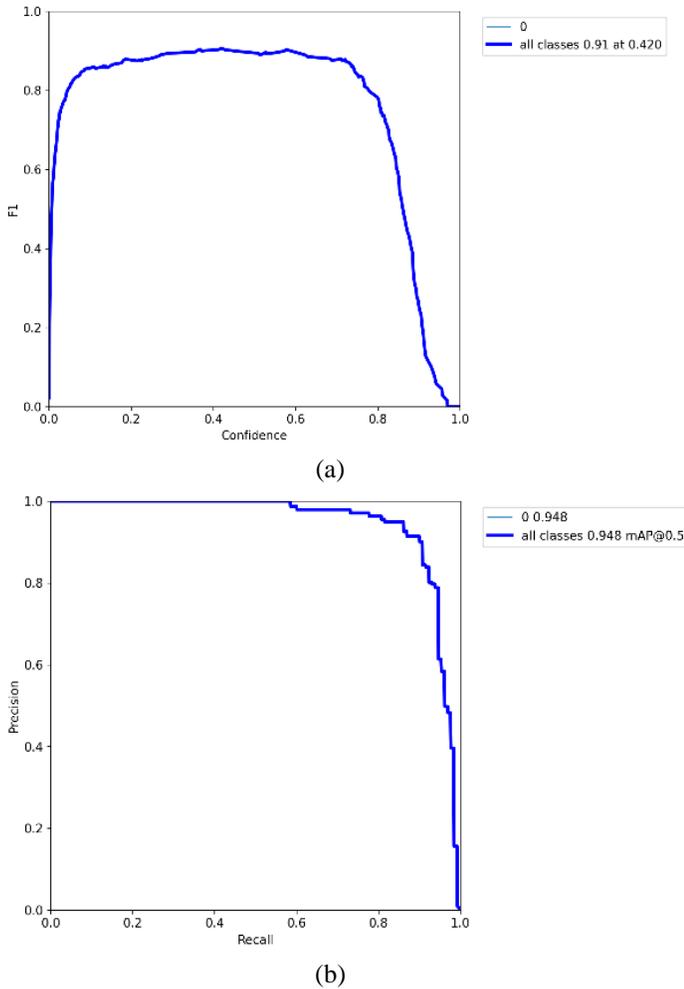

Figure 6. Test result.

Additionally, we train models using both the original data and the augmented data separately, and the test results are presented in Table 1. This demonstrates that data augmentation has improved the performance of the model.

Table 1. Test results on different datasets.

| Dataset | P (%) | R (%) | F1 score (%) | mAP (%) |
|---|---|---|---|---|
| Origin | 88.2 | 86.1 | 87.0 | 90.2 |
| Augmented | 91.4 | 90.0 | 91.0 | 94.8 |

## 5. CONCLUSION AND DISCUSSIONS

### 5.1. Summary

To address the alignment issue between the optical transmitters (light sources) and the receivers (solar cells) in OWPT systems, this study has employed deep learning methods to detect solar cell receivers. Initially, a dataset containing four types of solar cells collected in various environments was compiled, followed by data preprocessing and image enhancement. Subsequently, the YOLO series algorithm, emerging in the field of object detection, particularly the YOLOv5-lite model, known for its lighter, faster, and easier deployability, was chosen. After training, results such as F1 score and mAP were obtained, indicating that YOLOv5-Lite can be used swiftly and accurately for solar cell detection, demonstrating significant potentials.

Due to the small size of the dataset, the experimental results of this study contain limitations. However, the methodology employed herein is generalizable and adaptable to different types of solar cells and scenarios encountered in practical applications. Thus, this study has verified the feasibility of deep learning methods for solar cell detection, which can be used in many applications, e.g., OWPT systems, enriching research in this aspect.

### 5.2. Discussions

Through experiments, this paper has obtained results of solar cell recognition using the YOLOv5-Lite model. Although the results demonstrated an excellent performance, the lack of experimental results on solar cell recognition using non-deep learning methods prevents comparison between deep learning and traditional methods, thus not fully showcasing the superiority of deep learning models, including YOLOv5-Lite.

In addition, during the actual inference process, we encountered misidentification cases caused by complex scenarios, especially errors in recognizing targets with similar appearance features. To address this issue, we tried to utilize various methods. For instance, we established specific categories for similar items when annotating the training set to differentiate solar cells. Furthermore, considering that solar panels have wires on their surfaces for connecting small cells, we introduced images processed with edge detection into the training set to highlight this feature. However, despite these attempts, the final results did not show significant improvement, and in some cases, there was even a decrease in accuracy. This also becomes one of the directions for further research in the future.

## REFERENCES

[1] Putra A. W. Setiawan, Hirotaka Kato, and Takeo Maruyama, "Infrared LED marker for target recognition in indoor and outdoor applications of optical wireless power transmission system." *Japanese Journal of Applied Physics* **59** (2020): SOOD06.

[2] Tang Jing and Tomoyuki Miyamoto. "Target recognition function and beam direction control based on deep learning and PID control for optical wireless power transmission system." *2020 IEEE 9th Global Conference on Consumer Electronics (GCCE)*, Kobe, Japan, 2020, pp. 907-911.

[3] Sanders Michael and Jin S. Kang, "Utilization of polychromatic laser system for satellite power beaming." *2020 IEEE Aerospace Conference*. IEEE, 2020.

[4] Saha Anindo et al., "A wireless optical power system for medical implants using low power near-IR laser." *2017 39th Annual International Conference of the IEEE Engineering in Medicine and Biology Society (EMBC)*, Jeju, Korea (South), 2017, pp. 1978-1981.

[5] Nguyen Dinh Hoa, "Dynamic optical wireless power transfer for electric vehicles." *IEEE Access* **11** (2023): 2787-2795.

[6] Nguyen Dinh Hoa, "Optical wireless power transfer for moving objects as a life-support technology." *2020 IEEE 2nd Global Conference on Life Sciences and Technologies (LifeTech)*, Kyoto, Japan, 2020, pp. 405-408.

[7] Mohsan Syed Agha Hassnain, Haoze Qian, and Hussain Amjad. "A comprehensive review of optical wireless power transfer technology." *Frontiers of Information Technology & Electronic Engineering* **24**(6) (2023): 767-800.

[8] Kawashima Nobuki, Kazuya Takeda, and Kyoichi Yabe. "Application of the laser energy transmission technology to drive a small airplane." *Chinese Optics Letters* **5**(101) (2007): S109-S110.

[9] Wang Xin, Bodong Ruan, and Mingyu Lu. "Retro-directive beamforming versus retro-reflective beamforming with applications in wireless power transmission." *Progress In Electromagnetics Research* **157** (2016): 79-91.

[10] Lim Jaeyeong, Tariq Shamim Khwaja, and Jinyong Ha. "Wireless optical power transfer system by spatial wavelength division and distributed laser cavity resonance." *Optics Express* **27**(12) (2019): A924-A935.

[11] Sovetkin Evgenii and Ansgar Steland, "Automatic processing and solar cell detection in photovoltaic electroluminescence images." *Integrated Computer-Aided Engineering* **26**(2) (2019): 123-137.

[12] A. Kaoru, Kenta Moriyama, and Tomoyuki Miyamoto, "Preliminary Characterization of Robust Detection Method of Solar Cell Array for Optical Wireless Power Transmission with Differential Absorption Image Sensing." *Photonics* **9**(11) (2022): 861.

[13] Kamilaris Andreas, and Francesc X. Prenafeta-Boldú, "Deep learning in agriculture: A survey." *Computers and electronics in agriculture* **147** (2018): 70-90.

[14] Amara Jihen, Bassem Bouaziz, and Alsayed Algergawy. "A deep learning-based approach for banana leaf diseases classification." *Datenbanksysteme für Business, Technologie und Web - Workshopband* (2017): 79-88.

[15] Christiansen Peter et al., "DeepAnomaly: Combining background subtraction and deep learning for detecting obstacles and anomalies in an agricultural field." *Sensors* **16**(11) (2016): 1904.

[16] Sørensen René A. et al., "Thistle detection using convolutional neural networks." *EFITA WCCA 2017 Conference, Montpellier Supagro, Montpellier, France*. 2017, pp. 2-6.

[17] Yao Jia et al., "A real-time detection algorithm for Kiwifruit defects based on YOLOv5." *Electronics* **10**(14) (2021): 1711.

[18] Redmon Joseph et al., "You only look once: Unified, real-time object detection." *Proceedings of the IEEE conference on computer vision and pattern recognition*, Las Vegas, America, 2016, pp. 779-788.

[19] Wu Wentong et al., "Application of local fully Convolutional Neural Network combined with YOLO v5 algorithm in small target detection of remote sensing image." *PLoS One* **16**(10) (2021): e0259283.

[20] Gorban Alexander N., Evgeny M. Mirkes, and Ivan Y. Tyukin, "How deep should be the depth of convolutional neural networks: a backyard dog case study." *Cognitive Computation* **12**(2) (2020): 388-397.

[21] Tang Yi, "Exponential stability of pseudo almost periodic solutions for fuzzy cellular neural networks with time-varying delays." *Neural Processing Letters* **49**(2) (2019): 851-861.

[22] Ultralytics. Yolov5. Available online: https://github.com/ultralytics/yolov5

[23] Xiangrong Chen and Ziman Gong, YOLOv5-Lite: Lighter, faster and easier to deploy, 2021. Available online: https://github.com/ppogg/YOLOv5-Lite

[24] Roboflow. Available online: https://roboflow.com/

**Contact E-mail Address: hoa.nd@i2cner.kyushu-u.ac.jp**